\documentclass[aps,a4paper]{revtex4}
\usepackage{graphicx}

\newcommand{\be}{\begin{equation}}
\newcommand{\ee}{\end{equation}}
\newcommand{\ba}{\begin{array}}
\newcommand{\ea}{\end{array}}
\newcommand{\bea}{\begin{eqnarray}}
\newcommand{\eea}{\end{eqnarray}}
\newcommand{\bdm}{\begin{displaymath}}
\newcommand{\edm}{\end{displaymath}}
\newcommand{\bp}{\begin{picture}}
\newcommand{\ep}{\end{picture}}

\begin{document}

\title{Voter Dynamics on an Ising Ladder: Coarsening and Persistence} 
\author{Prabodh Shukla}%
\email{shukla@spht.saclay.cea.fr}
\affiliation{ Service de Physique Th$\acute{e}$orique \\ CEA Saclay, 91191 
Gif-sur-Yvette, France} 
\altaffiliation{ Permanent address: Physics Department, North 
Eastern Hill University, Shillong-793 022, India}


\begin{abstract}

Coarsening and persistence of Ising spins on a ladder is examined under
voter dynamics. The density of domain walls decreases algebraically with
time as $t^{-1/2}$ for sequential as well as parallel dynamics. The
persistence probability decreases as $t^{-\theta_s}$ under sequential
dynamics, and as $t^{-\theta_p}$ under parallel dynamics where $\theta_p
=2 \theta_s \approx .88$. Numerical values of the exponents are explained.
The results are compared with the voter model on one and two dimensional
lattices, as well as Ising model on a ladder under zero-temperature
Glauber dynamics. 

\end{abstract} 

\maketitle

\large{

\section{Introduction} 

The voter dynamics is a simple stochastic dynamics for Ising spins on a
lattice ~\cite{voter}. It does not require a Hamiltonian or a heat bath
for its specification. In the voter dynamics, the state of a spin at
(discrete) time $t+1$ is equal to the state of one of its nearest
neighbors (chosen randomly) at the previous time $t$. It is not obvious
if this dynamics should evolve a random configuration of spins into an
ordered configuration in the long time limit. However, analytic
solutions of the model show that an ordered state is reached on
lattices of dimensionality $d\le2$, but not on lattices of higher
dimensionality $d\ge3$ . Although a ferromagnetically ordered state of
spins is realised in $d=1$ as well as $d=2$, the approach to the
ordered state is qualitatively different in the two cases. For example,
the persistence probability decays algebraically as $t^{-\theta}$ in
$d=1$, but in $d=2$ it decays as $\exp[{- \mbox{constant}
(\log{t})^2}]$ ~\cite{persistence}.  The existence of apparently
universal features (independent of the initial configuration)
characterizing the approach to a steady state, and their likeness to
some non-equilibrium phenomena observed in nature has generated a good
deal of interest in the study of the voter dynamics. 

It is easy to understand the tendency of the voter dynamics to produce an
ordered state, if we imagine replacing the spins with balls of different
colors. Consider a large but finite lattice of $N$ points, and $N$ balls
of $N$ different colors. Initially, at $t=0$, each lattice point is
occupied by a ball whose color is unique to that lattice point. Now pick a
ball at random, repaint it with the color of one of its neighbors, and put
it back in its place. Repeat this process next time, and so on. At each
time, the number of colors on the lattice is reduced by one or zero. 
Therefore, after sufficiently long time the entire lattice will be
populated by balls of the same color. We may expect this tendency
irrespective of the dimensionality of the lattice as long as the lattice
is finite, and the number of time steps is sufficiently large. But the
limits $N \rightarrow \infty$ and $t \rightarrow \infty$ do not commute.
If the limit $N \rightarrow \infty$ is taken before $t \rightarrow
\infty$, a mapping of the voter model on to a system of coalescing random
walkers on a d-dimensional lattice brings out the d-dependent behavior
mentioned in the preceding paragraph.

On account of its simplicity, amenability to analysis, and rich behavior,
the voter dynamics has emerged as a paradigm for a broad class of
many-body stochastic systems. It has been studied extensively on regular
d-dimensional hypercubic lattices as well as networks and inhomogeneous
lattices ~\cite{castellano,suchecki,sood,mobilia}.  Variants of the voter
model ~\cite{variants} have been applied for modeling complex social
phenomena ~\cite{opinions} and some non-equilibrium physical phenomena
including spinodal decomposition~\cite{spinodal} and catalytic reactions
~\cite{catalysis}. Non-equilibrium physical phenomena belong to the realm
of thermal physics, and are usually studied in their simplest form by
using Glauber dynamics of Ising spins~\cite{glauber}. Similarity in the
behavior of a system under voter and Glauber's dynamics is surprising at
first sight. The Glauber dynamics is based on ideas of equilibrium thermal
physics. Glauber's spin-flip rates are chosen to satisfy the Boltzmann
distribution and detailed balance in thermal equilibrium. The same
transition rates are then exported outside the realm of equilibrium
statistical mechanics with the assumption that they would bring a
non-equilibrium system to equilibrium. On the other hand, concepts of a
Hamiltonian, Boltzmann's distribution, or detailed balance do not exist in
voter dynamics. The voter dynamics is based on ideas of random walks and
coarsening without surface tension. Yet, voter dynamics in $1d$ is
identical to the zero temperature Glauber dynamics of $1d$ Ising model. In
higher dimensions as well the two dynamical schemes often produce similar
caricatures of a system. It is interesting to speculate if these
similarities are fortuitous or perhaps the Glauber dynamics contains more
than what is minimally required to describe non-equilibrium phenomena such
as coarsening and persistence.

The present paper makes a comparative study of the voter dynamics and the
zero temperature Glauber dynamics of Ising spins on a ladder using
sequential as well as parallel dynamics. A ladder comprises two chains of
Ising spins placed next to each other in such a way that each spin has
three neighbors at equal distance. The ladder (visualized by imagining
bonds between nearest neighbors) is thus one step away from a $1d$ lattice
towards a $2d$ lattice. Spin ladders are used for modeling several
physical systems and are of current interest in the field of low
dimensional correlated electron systems~\cite{ladders}. However, our
primary motivation for studying voter dynamics on a ladder does not arise
from a direct interest in a physical system but rather from the fact that
a ladder is the simplest topology that distinguishes voter dynamics from
the zero temperature Glauber dynamics. In the same vein, we study parallel
as well as sequential dynamics. So far the voter model and its variants
have been studied mainly with sequential dynamics although social
processes such as voting are essentially parallel processes in the sense
that a large population votes together. It is therefore interesting to
examine if new features emerge in the behavior of a model under parallel
dynamics. In the case examined here, it turns out that the fraction of
voters who never change their vote up to time $t$ decreases much faster in
parallel dynamics than in sequential dynamics.

The zero temperature Glauber dynamics on a ladder is remarkably different
from one and two dimensional cases.  In $1d$ and $2d$, it shows coarsening
and algebraic decay of persistence with d-dependent persistence exponents.
There is no coarsening or algebraic decay of persistence on a ladder. The
reason is simple.  Coarsening and decay of persistence on a linear or
square lattice is related to the random motion of zero field sites, i.e.  
sites that have as many up neighbors as down. This is not possible on a
lattice with an odd coordination number. On a ladder, under zero
temperature Glauber dynamics, an initially random configuration of spins
gets stuck into a fixed point configuration of the type,

\bp(200,100)(-10,-10)

\thicklines

\multiput(20,50)(30,0){3}{\circle*{4}}
\multiput(20,20)(30,0){3}{\circle*{4}}

\multiput(100,10)(30,0){2}{\vector(0,1){20}}
\multiput(100,40)(30,0){2}{\vector(0,1){20}}

\multiput(160,30)(30,0){3}{\vector(0,-1){20}}
\multiput(160,60)(30,0){3}{\vector(0,-1){20}}

\multiput(250,10)(30,0){2}{\vector(0,1){20}}
\multiput(250,40)(30,0){2}{\vector(0,1){20}}

\multiput(310,30)(30,0){3}{\vector(0,-1){20}}
\multiput(310,60)(30,0){3}{\vector(0,-1){20}}

\multiput(400,50)(30,0){3}{\circle*{4}}
\multiput(400,20)(30,0){3}{\circle*{4}}

\ep \\

The two rows of arrows in the above figure represent segments of spins
along the two legs of the ladder. In a fixed point configuration the
nearest neighbor spins across the width of the ladder are aligned
parallel to each other, and there are ferromagnetic domains of various
lengths along the length of the ladder. The minimum length of a stable
domain is two lattice spacings. There is a huge degeneracy of fixed
point states. Therefore, under zero temperature sequential dynamics, an
initial random configuration quickly settles into a nearby fixed point.
Parallel dynamics gives qualitatively similar results but produces
limit cycles of period two in addition to fixed point states. An
example of a limit cycle of period two is given below. In this example,
spins labeled $s_{i}$ and $\sigma_{i}$ (with circles drawn around them
to aid the eye) flip at every time step while other spins remain fixed.
The labeling scheme and the diagonal lines are explained in the
following section. 

\bp(200,100)
\thicklines

\multiput(40,50)(30,0){2}{\circle*{4}}
\multiput(40,20)(30,0){2}{\circle*{4}}

\put(85,70){$\sigma_{i-3}$}
\put(135,70){ $s_{i-2}$ }
\put(185,70){$\sigma_{i-1}$}
\put(240,70){ $s_{i}$ }
\put(290,70){$\sigma_{i+1}$}
\put(340,70){ $s_{i+2}$ }
\put(390,70){ $\sigma_{i+3}$ }

\put(85,-5){$s_{i-3}$}
\put(135,-5){ $\sigma_{i-2}$ }
\put(185,-5){$s_{i-1}$}
\put(240,-5){ $\sigma_{i}$ }
\put(290,-5){$s_{i+1}$}
\put(340,-5){ $\sigma_{i+2}$ }
\put(390,-5){ $s_{i+3}$ }

\multiput(100,30)(50,0){3}{\vector(0,-1){20}}
\multiput(100,60)(50,0){3}{\vector(0,-1){20}}

\multiput(250,40)(30,0){1}{\vector(0,1){20}}
\multiput(250,50)(30,20){1}{\circle{25}}
\multiput(250,30)(30,0){1}{\vector(0,-1){20}}
\multiput(250,20)(30,20){1}{\circle{25}}

\multiput(300,10)(50,0){3}{\vector(0,1){20}}
\multiput(300,40)(50,0){3}{\vector(0,1){20}}

\multiput(430,50)(30,0){2}{\circle*{4}}
\multiput(430,20)(30,0){2}{\circle*{4}}

\thinlines
\multiput(105,60)(100,0){1}{\line(3,-4){35}}
\multiput(205,60)(100,0){1}{\line(3,-4){30}}
\multiput(305,60)(100,0){1}{\line(3,-4){38}}

\multiput(160,10)(100,0){1}{\line(3,4){35}}
\multiput(265,20)(100,0){1}{\line(3,4){30}}
\multiput(358,10)(100,0){1}{\line(3,4){35}}

\ep \\

The zero temperature fixed points or limit cycles become metastable at a
low temperature $T$ because spins on a domain wall can flip at a small
rate even if it increases their energy. The lifetime of a metastable state
is of the order of $exp(2J/k_BT)$ where $2J$ is the energy barrier on a
domain wall, and $k_BT$ is the thermal energy. There is no corresponding
metastability in the voter dynamics. 

\section{Coarsening}

We represent Ising spins on a ladder as a $2 \times N$ matrix labeled by
$S_{j,i}=\pm1$; $j=1,2$; $i=1,2, \ldots N$; and consider voter dynamics
that evolves a random initial configuration of spins $\{S_{j,i}(t=0)\}$
according to the following rule: 

\be S_{j,i}(t+1) = S_{j,i-1}(t) \mbox{ or } S_{k,i}(t) \mbox{ or }
S_{j,i+1}(t); \mbox{ each with probability } \frac{1}{3};\mbox{ } k=j+1
\mbox{ mod } 2 \ee

In other words, $S_{j,i}(t+1)$ takes one of three equally likely values;
$S_{j,i-1}(t)$, $S_{k,i}(t)$, or $S_{j,i+1}(t)$. Periodic boundary
conditions are imposed, i.e. $S_{j,i+N} = S_{j,i}$. The dynamics can be
implemented sequentially (updating one site at a time) or in parallel
(updating all sites simultaneously). We study both cases. There are
similarities as well as differences in the two cases. In both cases the
dynamics orders an initially disordered lattice. The rate at which
ordering occurs is comparable if time is appropriately rescaled so that N
steps of sequential dynamics correspond to a single step of parallel
dynamics. In the case of sequential dynamics, all spins get aligned
parallel to each other in the limit $t \rightarrow \infty$. In the case of
parallel dynamics, the lattice falls into two sub-lattices. Each
sub-lattice gets aligned ferromagnetically in the limit $t \rightarrow
\infty$, but the two sub-lattices may be aligned parallel or anti-parallel
to each other. In case they are aligned anti-parallel to each other, the
dynamics ends in a limit cycle of period two rather than a fixed point. 

In the process of reaching a fixed point or a limit cycle, the system
develops domains of up spins, down spins, and spins in an anti-
ferromagnetic arrangement.  As time progresses the domains become
larger and fewer in number in a self-similar way. In other words the
system coarsens ~\cite{bray}. In order to understand the coarsening
quantitatively, it is convenient to divide the ladder into two
sub-lattices at the beginning itself. It simplifies the notation, and
helps to discuss parallel dynamics with greater clarity. Thus in place
of $2 N$ spins $S_{j,i}$, we consider two sets of $N$ spins denoted by
$\{s_{i}\}$ and $\{\sigma_{i}\}$, ${i=1,2,\ldots,N}$.  The new spins
lie on two one-dimensional chains zig-zagging through each other such
that $S_{1,i}=s_{1}$, $S_{2,i}=\sigma_{i}$; $S_{1,i+1}=\sigma_{i+1}$;
$S_{2,i+1}=s_{i+1}$; and so on. This is shown schematically in the
second picture in the previous section. The diagonal lines are drawn to
aid the eye in visualizing one of the two sub-chains (the $\sigma$
chain). In the new notation, $\{s_{i}(t+1)\}$ are determined by
$\{\sigma_{i}(t)\}$; and $\{\sigma_{i}(t+1)\}$ by $\{s_{i}(t)\}$. 

It is convenient to define correlation functions $C_{n}^{s}(t)$ and
$C_{n}^{\sigma}(t)$ for spins separated by $n$ units on each sub-lattice
at time $t$. 

\be C_{n}^{s}(t) = \frac{1}{N} \sum_{i}< s_{i}(t) s_{i+n}(t) > ;  \mbox{
} C_{n}^{\sigma}(t) = \frac{1}{N} \sum_{i}< \sigma_{i}(t) \sigma_{i+n}(t)
> \ee

The angular brackets denote average over sufficiently large number of
initial random configurations $\{s_{i}(0)\}$ and $\{\sigma_{i}(0)\}$. The
two sub-lattices are statistically similar, and for most purposes one can
drop the superscripts on $C_{n}$. Thus, under parallel dynamics,
$C_{n}(t+1)$ is given by,

\be C_{n}(t)= \frac{1}{9} \left[ C_{n-2}(t-1) + 2 C_{n-1}(t-1) + 3
C_{n}(t-1) + 2 C_{n+1}(t-1) + C_{n+2}(t-1) \right] \ee

The first term on the right-hand-side of the above equation can be
understood as follows: $s_{i}(t)$ is equal to $\sigma_{i+1}(t-1)$ with
probability $1/3$, and $s_{i+n}(t)$ is equal to $\sigma_{i+n-1}(t-1)$
with probability $1/3$. Thus $<s_{i}(t)s_{i+n}(t)>$ is equal to
$<\sigma_{i+1}(t-1)\sigma_{i+n-1}(t-1)>$ or $C_{n-2}(t-1)$ with
probability $1/9$. Other terms in the equation are obtained similarly. 
Successive iterations of equation (3) yield $C_n(t)$ in terms of
correlations $C_{n-2t}(0)$, $C_{n-2t+1}(0)$, $C_{n-2t+2}(0)$, $\ldots$,
$C_{n+2t}(0)$ in the initial state. The equation is linear, and
therefore we can easily write a closed form expression for $C_n(t)$ in
terms of the initial correlations. Before we do so, let us note some
general features of the above equation. We note that $C_{n-2}^{*}$ =
$C_{n-1}^{*}$ = $C_{n}^{*}=C_{n+1}^{*}=C_{n+2}^{*}=1$ is a fixed point
solution of equation (2) in the limit $t \rightarrow \infty$. We also
note that, by definition, $C_{0}(t)=1$, for all t. Therefore, a reduced
correlation function, $c_{n}(t) = C_{0}(t)-C_{n}(t)$ also satisfies the
same equation as $C_{n}(t)$. 

\be c_{n}(t)= \frac{1}{9} \left[ c_{n-2}(t-1) + 2 c_{n-1}(t-1) + 3
c_{n}(t-1) + 2 c_{n+1}(t-1) + c_{n+2}(t-1) \right] \ee 

Equation (4) has a fixed point solution $c_{n-2}^{*}$ = $c_{n-1}^{*}$ =
$c_{n}^{*}$ = $c_{n+1}^{*}$ = $C_{n+2}^{*}=0$, and we shall use this
equation to study the approach to the fixed point. A closed form
expression for $c_{n}(t)$ in terms of correlations in the initial state
can be obtained by several methods. The easiest method is to write a few
iterations of the equation explicitly, i.e. write $c_{n}(t)$ explicitly in
terms of appropriate correlations at time $t-2$, $t-3$ etc, and then
generalize the formula by induction. We get,

\be c_{n}(t)= \frac{1}{9^t} \left[ \sum_{l,r,d}^{t=l+r+d} \left( \ba{c}
l+r+d \\ l,r,d \ea \right) c_{n-2l+2r}(0) \right]. \ee

In the above equation, the sum is over integers $l$, $r$, and $d$ such
that $l+r+d=t$, and the symbol appearing immediately after the summation
sign is a trinomial coefficient,

\be \left( \ba{c} l+r+d \\ l,r,d \ea \right) = \frac
{(l+r+d)!}{{l!}{r!}{d!}} \ee

Equations (5) and (6) have a simple geometrical interpretation that is
natural to voter dynamics on a ladder.  Imagine a backward walk in
discrete time. A walker starts at site $i$ at time $t$ and visits those
sites where the spin $s_{i}(t)$ has resided previously. After $t$ steps,
the walker arrives at the original address of $s_{i}(t)$ in the initial
configuration of the system. The original address with respect to site $i$
is specified by the number of left turns $l$, the number of right turns
$r$, and the number of straight steps $d$ taken by the walker in $l+r+d=t$
steps. We can replace the single walk tracing the history of $s_{i}(t)$ by
a set of directed random walks that always move one step closer to the
initial configuration but take a left turn or a right turn or a straight
step with equal probability.  There are a total of $3^{t}$ such walks that
can land on any site $i-t$ to $i+t$ in the initial configuration. The
trinomial coefficient in equation (5) gives the number of random walks
that land at a site corresponding to the set of integers $\{l,r,d\}$. A
path that returns to the starting site after $t$ steps is characterized by
an equal number of left and right turns, i.e. $l=r$. The number of walks
that return to the starting point are the largest in number,

\bea c_{n}(t) \approx \frac{1}{3^{t}} \sum_{0}^{l_{max}} \left( \ba{c} t
\\ l,l,t-2l \ea \right) c_{n}(0) = \frac{1}{3^{t}}
\sum_{0}^{l_{max}}\left( \ba{c} 2 l \\ l \ea \right) \left( \ba{c} t \\ 2l
\ea \right) c_{n}(0); \nonumber \\ l_{max}= \frac{t}{2} \mbox{ if t even
}, \frac{t-1}{2} \mbox{ if t odd}.\eea

The two binomial coefficients appearing above result from the
factorization of the trinomial coefficient, and their meaning is quite
clear. The first specifies the number of ways that one can make a walk of
$l$ steps to the left, and an equal number of steps to the right. The
second gives the number of ways one can take the remaining $t-2l$ steps as
straight steps in a total of $t$ steps. As may be expected, the product of
the two binomial coefficients is largest when $l=r=d=t/3$. Using
Stirling's approximation for the factorials of large numbers, we find that
the maximum value of the product scales as $1/t$ in the limit $t
\rightarrow \infty$. In the neighborhood of this maximum, the number of
terms that contribute significantly to the sum in equation (7) is of the
order of $t^{1/2}$. This follows from the fact that the distance covered
by straight steps in the walk of $t$ steps has a Gaussian distribution
with mean value equal to $\frac{1}{3}t$, and variance proportional to $t$.
Consequently the sum of the product of the two binomial coefficients
scales as $t^{-1/2}$. The correlation functions $c_n(t)$ approach zero as
$t^{-1/2}$ and $C_n(t)$ defined by equation (2) approach unity with the
same power law in the limit $t \rightarrow \infty$. The density of domain
walls (number of domain walls per site) decreases and coarsening increases
with the same power law. This is easily seen by defining the density of
domain walls $\rho(t)$ at time $t$ as,

\be \rho(t) = \frac{1}{2N} \left[ \sum_{i} \left\{ 3^{2} - \{s_{i-1}(t) +
s_{i}(t) + s_{i+1}(t)\}^{2} \right\} + \sum_{i} \left\{ 3^{2} -
\{\sigma_{i-1}(t) + \sigma_{i}(t) + \sigma_{i+1}(t)\}^{2} \right\} \right]
\ee

The above expression has been constructed so that a site whose nearest
neighbors are in the same state does not contribute to the density of
domain walls in the system. Note that the site in question need not be in
the same state as its neighbors. This definition takes into account
independent evolutions of the two sub-lattices under parallel dynamics and
treats the conclusion of the coarsening dynamics in a fixed point or a
limit cycle on equal footing. The density of domain walls is related to
correlation functions by the equation $ \rho(t) = 2 \left[ 2 c_{1}(t) +
c_{2}(t) \right]$, and vanishes as $t^{-1/2}$ in the limit $t \rightarrow
\infty$. Figure (1) shows $\log{\rho(t)}$ vs $\log{t}$ for parallel and
random sequential dynamics obtained by computer simulation of the voter
dynamics on systems of size $N=200$, and $N=2000$. In each case, we have
taken an average over $10^{4}$ different realizations of the initial
random configuration. A striking feature of figure (1) is that results of
parallel dynamics are indistinguishable from those of sequential dynamics.
This may be expected on ground that the stochastic equation (3) describes
each step of sequential dynamics as well, and we have rescaled time such
that $2N$ steps of sequential dynamics correspond to a single step of
parallel dynamics. As expected, the power law $t^{-1/2}$ fits coarsening
very well over a period of time that increases with the size of the
system. Figure (1) also shows that the average coarsening for $N=200$ and
$N=2000$ deviates from a power law and approaches zero at
$t=t_{max}\approx N^{2}$ in a self-similar way. This is due to the fact
that the evolution of the system and consequently coarsening stops when
the dynamics reaches a fixed point or a limit cycle, say at time
$t=t_{end}$. The upper limit on $t_{end}$ is approximately $N^{2}$, but it
has a broad distribution that depends on $N$ as well as the initial
configuration of the system. In the present paper, we do not go into the
distribution of $t_{end}$, and confine ourselves only to the region where
the average coarsening exhibits an algebraic power law.

Before leaving this section, we briefly present for comparison the
results for coarsening in the $1d$ Ising model under voter or
zero-temperature Glauber dynamics. In this case, there are two
simplifications: (a) there is only one chain instead of two zig-zagging
chains, and (b) the backward random walk can only take a step to the
right or to the left i.e. there are no straight steps. With these 
simplifications, the $1d$ version of equation (4) becomes,

\be c_{n}^{1d}(t)= \frac{1}{4} \left[ c_{n-2}^{1d}(t-1) + 2
c_{n}^{1d}(t-1) + c_{n+2}^{1d}(t-1) \right] \ee Equation (9) has the
solution, \be c_{n}^{1d}(t)= \frac{1}{4^t} \left[ \sum_{m=0}^{2t}
\left( \ba{c} 2t \\ m \ea \right) c_{n-2t+2m}^{1d}(0) \right]. \ee

As $t\rightarrow \infty$, the largest contribution on the
right-hand-side comes from the term $m=t$ (corresponding to an equal
number of left and right turns in the random walk ), and we get

\bea c_{n}^{1d}(t) \approx \frac{1}{2^{t}} \left( \ba{c} 2t \\ t \ea
\right) c_{n}^{1d}(0) \approx \frac{1}{\sqrt{\pi t}} c_{n}^{1d}(0) \eea

The quantity $c_{2}^{1d}(t)$ is a measure of the density of domain
walls (sites whose nearest neighbor spins are aligned anti-parallel to
each other) in $1d$, and as well known, it is seen to decay with the
power law $t^{-1/2}$.

\section{Persistence}

Persistence and coarsening often coexist in a non-equilibrium system
~\cite{satya}. We therefore examine our system for persistence as well.
The persistence probability $P(t)$ is the probability that a randomly
picked site in the system has never flipped from its initial state up to
time $t$. Figure (2) shows a plot of $\log{P(t)}$ versus $\log{t}$
obtained from computer simulations of the system. The data for figure (2)
and figure (1) are taken from the same set of simulations. However, at
first sight, a power law fit to persistence does not appear to be of the
same quality as for coarsening. There is a regime after the initial
transient period that fits a power law $t^{-\theta_{s}}$ for sequential
dynamics, and $t^{-\theta_{p}}$ for parallel dynamics but this regime is
smaller than the corresponding regime where coarsening shows a power law.
It is not very clear why the region of algebraic persistence should be
noticeably shorter than it is for average coarsening. We shall discuss it
further below but we first look at the features of figure (2) that are
relatively easy to understand. 

On the basis of our simulations, we estimate $\theta_{p}=2\theta_{s}
\approx .88$. The relationship $\theta_{p}=2\theta_{s}$ is easy to
understand. Its origin lies in the fact that the ladder can be partitioned
into two sub-lattices $A$ and $B$ that evolve similarly but independently
of each other. At time $t=0$, $A(0)$ and $B(0)$ are uncorrelated.
Thereafter $A(t+1)$ is determined by $B(t)$ and $B(t+1)$ by $A(t)$.
Persistence of a site on the lattice $A+B$ at odd times $(t=1,3,\ldots )$
is independent of its persistence at even times $(t=2,4,\ldots )$. At odd
times and similarly at even times, the persistence is characterized by an
exponent appropriate for sequential dynamics on a sub-lattice. Although
spins on a sub-lattice are updated in parallel, the motion of domain walls
is effectively the same as it would be under random sequential dynamics.
The reason is that the order in which domain walls are relaxed does not
matter under random sequential dynamics. A sequential relaxation process
in which each domain wall on a sub-lattice is relaxed once would produce a
qualitatively similar state as obtained in a single step of parallel
relaxation. Therefore the probability that a given site on the ladder has
been persistent at odd times scales as $t^{-\theta_{s}}$, and similarly
the probability that it has been persistent at even times scales as
$t^{-\theta_{s}}$. The probability that a site has been persistent at odd
as well as even times, i.e. it has been persistent under parallel dynamics
on the lattice $A+B$ scales as $t^{-2\theta_{s}}$ giving
$\theta_{p}=2\theta_{s}$. A similar effect is observed in one dimensional
Ising model under the zero temperature parallel Glauber dynamics
~\cite{menon}.

We give a heuristic argument for the observed value of the persistent
exponent $\theta_{p}$ that yields $\theta_{p}=8/9$ in close agreement with
the numerical value. A similar argument for the $1d$ Ising model yields
the exact value of the exponent $\theta_{p}^{1d}=3/4$ in that case. It is
therefore possible that $\theta_{p}=8/9$ may be the exact value of the
exponent for the ladder (in the time regime indicated below, and with the
provision for a crossover behavior at later times), but we are not in a
position to assert this on the basis of our heuristic argument. It is
desirable to calculate the persistence exponent for the ladder using a
method similar to the one used in the case of the $1d$
chain~\cite{derrida}, but this is outside the scope of the present paper.
We note that the persistence exponent in parallel dynamics is larger than
the coarsening exponent. The typical size of a coarsening domain is of the
order of $t^{1/2}$, and typical separation between two persistent sites is
of the order of $t^{\theta_{p}}$. If $\theta_{p}> 1/2$, persistent sites
are separated by a distance much larger than a domain. Therefore,
correlations between persistent sites may be neglected. In other words,
the effect of different persistent sites on each other may be neglected
and one can focus on the stability of a single persistent site under the
random motion of domain walls in the system. Why is the separation between
persistent sites much larger in parallel dynamics than it is in sequential
dynamics? The reason lies in the bi-partite nature of the lattice.  The
two sub-lattices $\{s_{i}\}$ and $\{\sigma_{i}\}$ evolve similarly but
independently of each other. Each sub-lattice develops ferromagnetically
ordered domains due to coarsening. However, domains on the two
sub-lattices may be aligned parallel or anti-parallel to each other. A
region where the domains on the two sub-lattices overlap anti-parallel to
each other contributes zero to persistence because spins in this region
flip at every time step. This not only increases separation between
persistent sites but modifies their spatial distribution as well.
Persistent sites in sequential dynamics tend to cluster together but in
parallel dynamics they are scattered more uniformly.

We use the idea of a backward random walk mentioned after equation (6) to
relate the persistence probability at time $\tau+1$ to persistence
probability at an earlier time $\tau$. The reason for using a new label
$\tau$ in place of $t$ is as follows. We regard a long random walk of $t$
steps as a series of $3^{\tau}$ short walks of $\tau$ steps each, and
replace the average of a quantity over the long walk by the average of the
same quantity over $3^{\tau}$ realizations of the short walk of length
$\tau$. After this averaging process, appropriate results for the long
walk of $t$ steps are obtained by the transformation $\tau=
\log_{3}{(t/{\tau})}$, or $\tau \approx \log_{3}{t}$ for $t>>\tau$. Let us
focus on two nearest neighbor spins, say $\sigma_{i}$ and $s_{i}$
connected by a rung on the ladder. The spins $\sigma_{i}$ and $s_{i}$ lie
on different sub-lattices. At any time $\tau$, $\sigma_{i}(\tau)$ and
$s_{i}(\tau)$ are independent of each other. Let $P(\tau)$ be the
probability that $\sigma_{i}(\tau)$ and $s_{i}(\tau)$ are persistent up to
$\tau$ under parallel dynamics. This implies that $\sigma_{i}(\tau)$ and
$s_{i}(\tau)$ are aligned parallel to each other. As a result of
coarsening, $\sigma_{i}(\tau)$ and $s_{i}(\tau)$ are likely to be inside a
ferromagnetically ordered domain on their respective sub-lattices.  If the
ferromagnetic domains on the two chains are not aligned parallel to each
other, for example if
$\sigma_{i-1}(\tau-1)=\sigma_{i}(\tau-1)=\sigma_{i+1}(\tau-1)=-1$, and
$s_{i-1}(\tau-1)=s_{i}(\tau-1)=s_{i+1}(\tau-1)=1$, then we would get
$s_{i}(\tau)=-1$ and $\sigma_{i}(\tau)=1$ contrary to our assumption that
$\sigma_{i}(\tau)$ and $s_{i}(\tau)$ have never flipped up to $\tau$. Let
us focus on a specific case, say $\sigma_{i}(\tau) = s_{i}(\tau)=1$. We
now ask for the probability that $\sigma_{i}(\tau+1) = s_{i}(\tau+1)=1$,
or equivalently the probability $P(\tau+1)$ that $\sigma_{i}(\tau+1)$ and
$s_{i}(\tau+1)$ are persistent up to $\tau +1$ given that they are
persistent up to $\tau$. The value of $\sigma_{i}(\tau+1)$ is determined
by $s_{i-1}(\tau)$, $s_{i}(\tau)$, and $s_{i+1}(\tau)$. If $s_{i}(\tau)=1$
(with probability $1/2$), the probability that $\sigma_{i}(\tau+1)=1$ is
to be calculated over four configurations of the spins $s_{i-1}(\tau)$ and
$s_{i+1}(\tau)$. In calculating the average over configurations, each
configuration occurs with equal weight but contributes differently to
$\sigma_{i}(\tau+1)$. Keeping in mind the rules of the voter dynamics and
$s_{i}(\tau)=1$, the configuration $\{s_{i-1}(\tau)=1,s_{i+1}(\tau)=1\}$
gives $\sigma_{i}(\tau+1)=1$ with probability unity. The configurations
$\{s_{i-1}(\tau)=1,s_{i+1}(\tau)=-1\}$ and
$\{s_{i-1}(\tau)=-1,s_{i+1}(\tau)=1\}$ each give $\sigma_{i}(\tau+1)=1$
with probability $2/3$. Finally the configuration
$\{s_{i-1}(\tau)=-1,s_{i+1}(\tau)=-1\}$ gives $\sigma_{i}(\tau+1)=1$ with
probability $1/3$. Thus the probability to get $\sigma_{i}(\tau+1)=1$
equals $(1+2/3+2/3+1/3)/4$ or $2/3$. Similarly, the probability to get
$s_{i}(\tau+1)=1$ is also equal to $2/3$. Putting various probabilities
together, we get

\be P(\tau+1) = \frac{1}{4} \left (\frac{2}{3}\right)^{2} P(\tau) ;  
\mbox{ or } P(\tau+1)-P(\tau) = -\frac{8}{9} P(\tau) \ee

Treating $\tau$ as a continuous variable in the limit $\tau \rightarrow 
\infty$ and making the identification $\tau=\log{t}$, we get 

\be \left. \frac{dP(\tau)}{d\tau}= -\frac{8}{9} P(\tau);\mbox{ or }
P(\tau)=\exp -\frac{8}{9} \tau ; \mbox{ or } P(t) \approx t^{-8/9}
\right.\ee

A similar argument yields the exact value of the persistence exponent
$\theta_{p}^{1d}=3/4$ in the case of $1d$ Ising model evolving under voter
or zero temperature Glauber dynamics. As mentioned earlier, there are two
simplifications in the $1d$ case. We have only a single chain of spins
$s_{i}(\tau)$, and the backward random walk can only take a step to the
left or to the right. Under parallel dynamics $s_{i}(\tau+1)$ is
determined by $s_{i-1}(\tau)$ and $s_{i+1}(\tau)$. Therefore
$s_{i}(\tau+1)$ is independent of $s_{i}(\tau)$. Let $P^{1d}(\tau)$ be the
probability that a site is persistent up to $\tau$. A persistent site could
be in the state $s_{i}(\tau)=-1$ or $s_{i}(\tau)=1$ with probability equal
to $P^{1d}(\tau)/2$ . Let us suppose $s_{i}(\tau)=1$, and ask for the
probability that $s_{i}(\tau+1)=1$ as well. The configuration
$\{s_{i-1}(\tau)=1,s_{i+1}(\tau)=1\}$ gives $s_{i}(\tau+1)=1$ with
probability unity. The configurations
$\{s_{i-1}(\tau)=1,s_{i+1}(\tau)=-1\}$ and
$\{s_{i-1}(\tau)=-1,s_{i+1}(\tau)=1\}$ each give $s_{i}(\tau+1)=1$ with
probability $1/2$. The configuration
$\{s_{i-1}(\tau)=-1,s_{i+1}(\tau)=-1\}$ gives $s_{i}(\tau+1)=1$ with
probability zero. Therefore the probability to get $s_{i}(\tau+1)=1$
equals $(1+1/2+1/2+0)/4$ or $1/2$. We get,

\be P^{1d}(\tau+1) = \frac{1}{4}P^{1d}(\tau) ;  \mbox{ or }
P^{1d}(\tau+1)-P^{1d}(\tau) = -\frac{3}{4} P^{1d}(\tau); \mbox{ or }
P^{1d}(t) \approx t^{\theta_p^{1d}} \approx
 t^{-3/4} \ee

The value $\theta_p^{1d}=3/4$ corresponds to the exact result for the
persistence exponent $\theta_s^{1d} = 3/8$ for the one dimensional Ising
model under sequential dynamics~\cite{derrida}. However, the derivation of
the exact result is based on the limit $t \rightarrow \infty$ in a finite
system with periodic boundary conditions. Our equation relating
$P^{1d}(\tau)$ to $P^{1d}(\tau+1)$ under parallel dynamics is intended
over time scales $\log{t}<< N$. The number of backward time steps $\tau$
should be large but not so large that backward walks from each persistent
site end in overlapping regions in the initial state of the system. Thus
the agreement of our result with the exact answer for Ising spins may be
fortuitous but it is interesting in view of the difficulty of the exact
calculation. We do not expect the same argument to reproduce the exact
values of the persistence exponent for the one dimensional $q-$state Potts
model for $q>2$ ~\cite{derrida}. In this case the exact values of the
exponents correspond to distances between persistent sites under parallel
dynamics that are larger than the length of the chain.

\section{Conclusion}

The voter dynamics on an Ising ladder exhibits coarsening and persistence
that is qualitatively similar to the one observed in the one dimensional
Ising model under the zero-temperature Glauber dynamics. In both cases,
coarsening scales as $t^{1/2}$ in the limit $t \rightarrow \infty$,
although the technical details are somewhat different. This behavior is
easily understood in view of the diffusive nature of coarsening. For a
finite system of $N$ spins, the longest time over which coarsening is
observed scales as $N^{2}$. However, in a typical numerical simulation of
the system, a fully ordered state may be reached in a time much less than
the upper limit mentioned above. The time taken by the dynamics to put the
system into a completely ordered state and thus to come to a stage where
it stops evolving depends significantly on the initial configuration of
the system. It is not clear exactly what features of the initial random
configuration are responsible for a broad spectrum of life times of
coarsening. Given an initial random configuration, we are not in a
position to predict how long it would take the dynamics to reach a fixed
point or a limit cycle. Numerical simulations show that coarsening when
averaged over a sufficiently large number of initial random configurations
of the system begins to deviate from the algebraic law $t^{1/2}$ several
decades before reaching $t_{end} \approx N^{2}$. 

Persistence also shows an algebraic power law $t^{-\theta}$, but when
compared with coarsening in numerical simulations, the power law for
persistence is observed over a smaller regime. Unlike the coarsening
exponent, the persistent exponent $\theta$ depends on what type of
dynamics is used, parallel or sequential. The exponent for the parallel
dynamics $\theta_{p}$ is twice the exponent $\theta_{s}$ for sequential
dynamics. The relationship $\theta_{p}=2\theta_{s}$ is easily understood
in view of the bi-partite nature of the ladder lattice. We have given a
heuristic argument to suggest $\theta_{p}=8/9$, a value that is close to
the numerically observed value. Numerical simulations show considerable
persistence in the system even after the exponent deviates from the value
mentioned above. It is possible that in the limit $t \rightarrow \infty$,
the persistent exponent crosses over to the corresponding exponent for the
one dimensional Ising model under the zero temperature Glauber dynamics.
If this were so, it would mean that the voter dynamics on a ladder is in
the same universality class of non-equilibrium phenomena as the one
dimensional Ising model under Glauber dynamics. However, this issue is
difficult to decide numerically because it is computationally intensive.
It already takes several days to generate the data shown in figures (1)
and (2). We expect that with increasing size of the ladder the exponents
$\theta_{p}=8/9$, and $\theta_{s}=4/9$ will be observed over longer
decades of time. We have suggested an explanation for the observed values
of the exponents over time scale $\log{t} <<N$. Numerical results leave
open the possibility that at longer times the persistence exponent under
voter dynamics on a ladder may cross over to the persistence exponent in
the one dimensional Ising model under zero temperature Glauber dynamics.

I thank C De Dominicis, C Godr$\acute{e}$che, and J M Luck for helpful
discussions.

\begin{figure}

\begin{center}

\includegraphics[angle=-90,width=16cm]{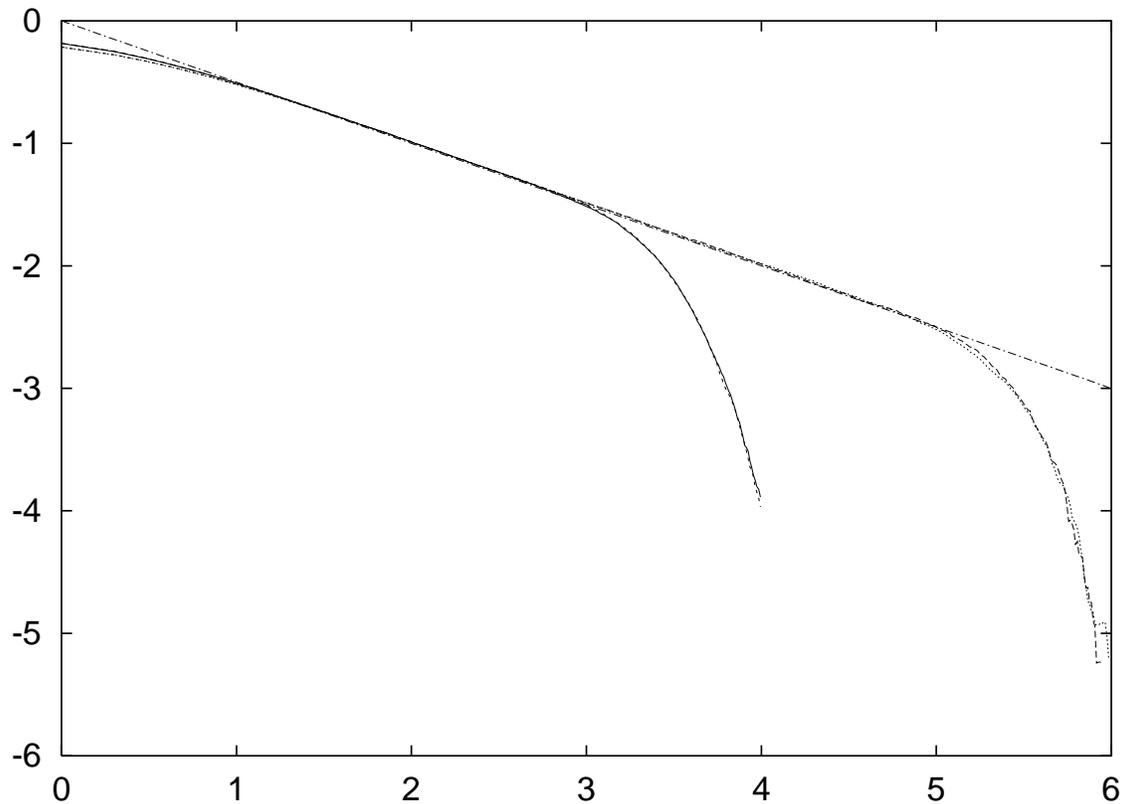}

\caption{ Coarsening: $\log {\rho(t)}$ versus $\log{t}$ for $N=2
\times 10^2$ (continuous line) and $N=2 \times 10^3$ (broken line)
averaged over $10^4$ realizations of initial random configuration.
Graphs for sequential and parallel dynamics are superposed on each
other and are nearly indistinguishable from each other. A line with
slope $-1/2$ has been drawn for comparison.}

\end{center}

\end{figure}

\begin{figure}

\begin{center}

\includegraphics[angle=-90,width=16cm]{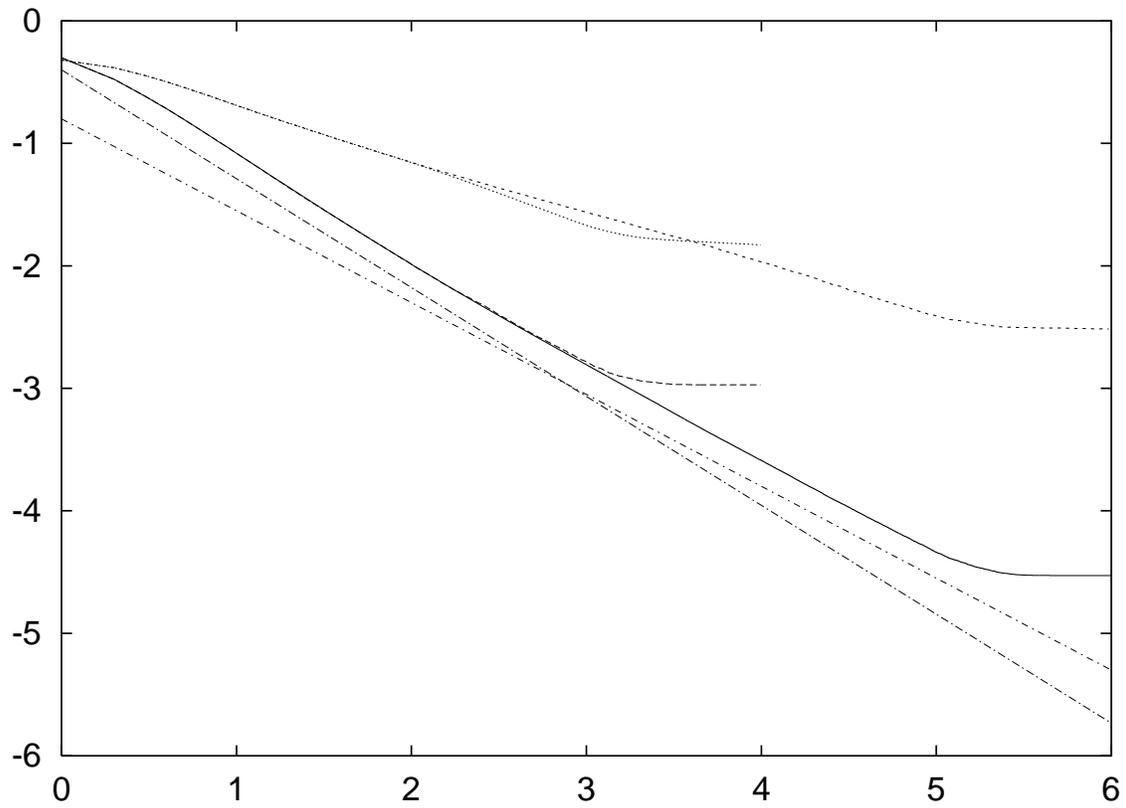}

\caption{ Persistence: $\log {P(t)}$ versus $\log{t}$ obtained from the
same computer simulations as used in figure (1). The lower curves are for
parallel dynamics and have approximately twice the slope of upper curves
for random sequential dynamics. A single power law does not appear to fit
persistence as well as it does coarsening. Two straight lines with slopes
$3/4$ and $8/9$ are drawn for comparison with parallel dynamics,
suggesting a crossover in the power law for persistence.}

\end{center}

\end{figure}

\end{document}